\documentclass[nature,superscriptaddress,twocolumn]{revtex4-1}

\usepackage[utf8]{inputenc}
\usepackage[english]{babel}
\usepackage{amsmath}
\usepackage{amsfonts}
\usepackage{amssymb}
\usepackage{graphicx}
\usepackage{amsthm}
\usepackage{amsbsy}
\usepackage[monochrome]{color}
\usepackage{hyperref}
\usepackage[caption=false]{subfig}
\newcommand{\ket}[1]{\ensuremath{|#1\rangle}}

\newcommand{\ketbra}[1]{\ensuremath{| #1 \rangle\langle #1 |}}

\newcommand{\eqq}[1]{Eq.~(\ref{#1})}
\newcommand{\tr}[2]{\ensuremath{\operatorname{Tr}_{\mathrm{#1}}\left[#2\right]}}
\newcommand{\abs}[1]{\ensuremath{\left|#1\right|}}
\newcommand{\en}{\ensuremath{N}}
\newcommand{\vecc}[1]{\ensuremath{\pmb{#1}}}

\begin{document}
\title{Narrow Bounds for the Quantum Capacity of Thermal Attenuators} 
\author{Matteo Rosati}
\email{matteo.rosati@uab.cat}
\affiliation{F\'{\i}sica Te\`{o}rica: Informaci\'{o} i Fen\`{o}mens Qu\`{a}ntics, Departament de F\'{\i}sica, Universitat Aut\`{o}noma de Barcelona, 08193 Bellaterra, Spain}
\affiliation{NEST, Scuola Normale Superiore and Istituto Nanoscienze-CNR, I-56127 Pisa, Italy.}
\author{Andrea Mari} 
\affiliation{NEST, Scuola Normale Superiore and Istituto Nanoscienze-CNR, I-56127 Pisa, Italy.}
\author{Vittorio Giovannetti} 
\affiliation{NEST, Scuola Normale Superiore and Istituto Nanoscienze-CNR, I-56127 Pisa, Italy.}

\maketitle

\section*{Abstract}
{\color{red}Thermal attenuator channels model the decoherence of quantum systems interacting with a thermal bath, e.g., a two-level system subject to thermal noise and an electromagnetic signal travelling through a fiber or in free-space. Hence determining the quantum capacity of these channels is an outstanding open problem for quantum computation and communication. 
Here we derive several upper bounds on the quantum capacity of qubit and bosonic thermal attenuators.
We introduce an extended version of such channels which is degradable and hence has a single-letter quantum capacity, bounding that of the original thermal attenuators. Another bound for bosonic attenuators is given by the bottleneck inequality applied to a particular channel decomposition. With respect to previously known bounds we report better results in a broad range of attenuation and noise: we can now approximate the quantum capacity up to a negligible uncertainty for most practical applications, e.g., for low thermal noise.}
%In this article we derive several upper bounds on the quantum capacity of qubit and bosonic thermal attenuators, commonly employed to model the effect of loss and thermal noise in real physical systems and communication processes. 
%We introduce an extended version of such channels which is degradable and hence has a single-letter quantum capacity; the latter is an upper bound for the capacity of the original thermal attenuators that is tight in the limit of low thermal noise. For bosonic attenuators we also introduce a further bound based on the bottleneck inequality applied to a particular channel decomposition.   With respect to previously known bounds we report better results in a broad range of attenuation and noise values, both for qubit and bosonic attenuators. Even if the exact quantum capacity of thermal attenuators is still unknown, we can approximate it up to an uncertainty which is negligible for most practical applications.

\section*{Introduction}
The study of information transmission between distant parties has attracted much theoretical attention since the seminal work of Shannon~\cite{shannonSeminal,shannonSeminal1}, who gave birth to the field of information theory by determining the ultimate limits for compression and transmission rate. The latter is called information capacity and it depends on the specific channel that is used to model a physical transmission process. Hence, since the information carriers are ultimately governed by the laws of quantum physics, in more recent years there has been interest in analyzing the communication problem in a quantum setting, giving birth to the field of quantum communication and information theory~\cite{holevoBOOK,cavesDrum,wildeBOOK,hayashiBOOK,nChuangBOOK}. Several results have been obtained so far, such as: an expression for the capacity of a quantum channel for the transmission of classical information~\cite{holevo1,holevo2,holevo3,schumawest1,schumawest2} and its explicit value for some classes of channels~\cite{wildeBOOK,hayashiBOOK,kingUnital,depolarizingCCap,gaussOpt,gaussMaj,gaussCap,LOSSY,amplitudeDamping}; the use of entanglement~\cite{horodecki}  as a resource for communication~\cite{bennetEntAss1,bennetEntAss2}; %an expression of 
the capacity of a quantum channel for the transmission of quantum information, i.e., of states preserving quantum coherence~\cite{lloydQuant,shorQuantum,qCap1,qCap2}.
The latter is called quantum capacity of the channel and constitutes the main focus of this work. As usual in information theory, although a formula exists for this quantity, it is quite difficult to compute for general channels, due for example to the necessary regularization that takes into account the use of entangled inputs over multiple channel uses~\cite{holGiovRev}. The problem simplifies considerably for the so called {\it degradable} channels ~\cite{deg1,antiDeg,deg2,amplitudeDamping,weakDeg}, but it also exhibits some striking features in other cases~\cite{DiVincenzo1998,Smith2007,zeroQCap,hastings2009,supAddQCap}.
An important class of channels is that of Thermal Attenuators (TAs), which describe the effect of energy loss due to the interaction with a thermal environment. Common examples of this class are the qubit thermal attenuator or  generalized amplitude damping channel~\cite{nChuangBOOK,amplitudeDamping} and its infinite-dimensional counterpart given by the bosonic Gaussian thermal attenuator \cite{cavesDrum, holevoBOOK, RevGauss}. Physically, the former is a good model for the thermalization process of a two-level system (e.g. a single-qubit quantum memory) in contact with a thermal bath, while the latter is the standard description for optical-fiber and free-space communications in the presence of thermal noise. {\color{red} Notice that thermal noise at room temperature is not negligible for low-frequency electromagnetic signals like, e.g., infra-red lasers, microwaves, radio waves, {\it etc.}. For example, a crude estimate of the thermal noise in a real communication channel can be obtained by using  the Bose-Einstein distribution at the desired wavelength to estimate the excess contribution to the mean photon number of the transmitted signal: this grows from $O(10^{-14})$ at telecom wavelengths, i.e., $1550 nm$, and $O(10)$ at microwave wavelengths, i.e., $1 mm$ and above. Accordingly, although
the thermal noise may be negligible at telecom wavelengths, it becomes increasingly important as one spans the whole optical domain and it is a crucial parameter in the microwave domain. Hence the study of information transmission on TAs is of particular relevance for future quantum communication networks where hybrid interfaces will be employed, e.g., superconducting qubits connected by
microwave communication lines, see Refs.~\cite{Xiang2017,Kurizki2015}.}
In both the qubit and bosonic cases, the corresponding quantum capacity is known \cite{amplitudeDamping,qCapLossy} only for a zero-temperature environment, since in this limit both channels are degradable~\cite{deg1,antiDeg,deg2}.
However the more general finite-temperature case breaks the degradability property and it has not been successfully tackled so far, apart from establishing some upper and lower bounds \cite{holWer,plob,wildeDecDirettaBound}.
 
 In this article we introduce a general method to compute new bounds on the capacity of any quantum channel which is {\it weakly degradable} in the sense of~\cite{weakDeg,antiDeg}. It is based on the purification of the environment by an additional system, which is then included in the output of an extended version of the original channel. Degradability is restored for such extended channel and its quantum capacity can be easily computed, providing an upper bound on the capacity of the original channel, which is tight in the limit of small environmental noise. This result applies to any weakly degradable channel but, in order to be more explicit,  here we compute the associated upper bound for both the qubit and continuous-variable thermal attenuators. Moreover, for the Gaussian attenuator, we provide an additional bound based on the bottleneck inequality applied to a  twisted version of the channel decomposition used in~\cite{deco2,weakDeg,gaussOpt}. Eventually we compare all our bounds with those already known in the literature and in particular with the recent results of~\cite{plob,wildeDecDirettaBound}. We show a significant improvement with respect to the state-of-the-art for a large region of attenuation and noise  parameters,  shrinking the unknown value of the quantum capacity within an error bar which is so narrow to be irrelevant for many practical applications.   \\

%The article is structured as follows: in Sec.~\ref{sec:thAttExt} we briefly review the class of thermal attenuator channels and their quantum capacity problem; in Sec.~\ref{sec:extCh} we introduce the extended channel, and analyze its properties in relation to the original channel and finally provide the expressions of the upper bounds; in Sec.~\ref{sec:altBound} we introduce a further bound for the bosonic case; eventually, in Sec.~\ref{sec:conc} we draw our conclusions.  Technical details are left for the Methods Section.

\section*{Results}

\subsection*{Thermal Attenuator channels}\label{sec:thAttExt}
\begin{figure}[t!]
\includegraphics[scale=.37,trim={5.5cm 2.5cm 7.5cm 2.5cm},clip]{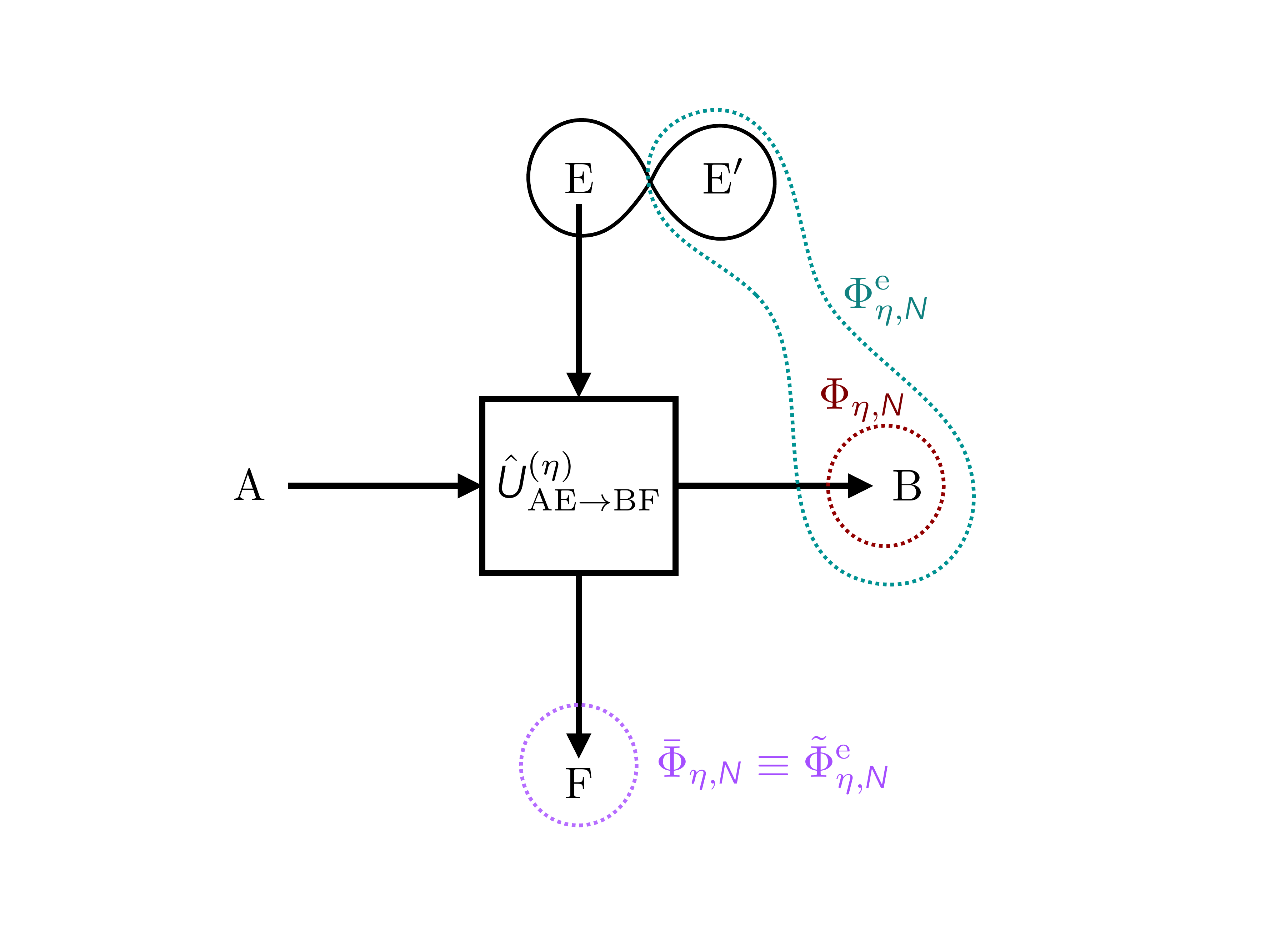}
{\color{red}\caption{Unitary representation of a thermal attenuator channel, its extended channel and their complementaries.\\ The system A interacts with the environment E, which is in a thermal state of mean energy $N$, via a linear coupling, $\hat{U}^{(\eta)}_{\mathrm{AE}\rightarrow \mathrm{BF}}$. The parameter $\eta$ determines the fraction of energy dispersed from the system into the environment, while the total energy is preserved. The output of the channel $\Phi_{\eta,N}$ is recovered by discarding the output environment F, while that of its weak complementary $\bar{\Phi}_{\eta,N}$ by discarding the output system B. The two channels are called weakly complementary to each other because E is in a mixed state and hence this unitary representation is not a proper Stinespring dilation. By purifying the input environment via an ancilla E', entangled with E, we obtain the extended channel $\Phi_{\eta,N}^\mathrm{e}$ with output BE' and its strong complementary $\tilde{\Phi}_{\eta,N}^\mathrm{e}$ with output F. Note that the latter coincides with $\bar{\Phi}_{\eta,N}$. }\label{fg1}}
\end{figure}

Let us consider a communication channel that connects two parties. The sender, Alice, wants to transmit the quantum state $\hat{\rho}_{\mathrm{A}}\in\mathfrak{S}(\mathcal{H}_{\mathrm{A}})$, represented by a positive density operator of unit trace on the Hilbert space of the system, $\mathcal{H}_{\mathrm{A}}$. At the other end of the channel Bob receives a transformed quantum state on the Hilbert space $\mathcal{H}_{\mathrm{B}}$. Any physical transformation applied to the state during the transmission can be represented by a quantum channel $\Phi$, i.e., a linear, completely-positive and trace-preserving map, which evolves the initial state as $\hat{\rho}_{\mathrm{B}}=\Phi\left(\hat{\rho}_{\mathrm{A}}\right)$. \\
The class of channels studied in this article is that of thermal attenuators, $\Phi_{\eta,\en}$, originating from the following physical representation (see Fig.~\ref{fg1}). An energy-preserving interaction $\hat{U}^{(\eta)}_{\mathrm{AE}\rightarrow \mathrm{BF}}$, parametrized by a transmissivity parameter $\eta\in[0,1]$, couples the input system $\hat{\rho}_{\mathrm{A}}$ with an environment described by a thermal state $\hat{\tau}_{\mathrm{E}}\propto \exp(-\hat{H}_\mathrm{E})$, where $\hat{H}_\mathrm{E}$ is the bath hamiltonian, of dimension equal to that of the system, and the state has mean energy $N=\tr{}{\hat{H}_\mathrm{E}\hat{\tau}_\mathrm{E}} \geq0$ in dimensionless units. The total output state can be written as
\begin{equation}\label{phiDil}
\hat{\rho}_{\mathrm{BF}}=\hat{U}^{(\eta)}_{\mathrm{AE}\rightarrow \mathrm{BF}}\left(\hat{\rho}_{\mathrm{A}}\otimes\hat{\tau}_{\mathrm{E}}\right)\hat{U}_{\mathrm{AE}\rightarrow \mathrm{BF}}^{(\eta)\dag},
\end{equation}
while the action of the channel is obtained by tracing out the output environmental system F: 
\begin{equation}\label{generalTA}
\Phi_{\eta,\en}\left(\hat{\rho}_{\mathrm{A}}\right)=\tr{F}{\hat{\rho}_{\mathrm{BF}}}.
\end{equation}
This general framework is particularly relevant in the two paradigmatic cases in which the system is represented by a two-level system or by a single bosonic mode.  Both situations are quite important for practical applications since they model the effect of damping and thermal noise on common information carriers, typically used in experimental implementations of quantum information and quantum communication protocols. In the following we introduce in detail these two kinds of physical systems. 
\\

In order to describe a two-level system, we fix as a basis $\ket{0}$ and $\ket{1}$ corresponding to the ground and excited states respectively. In this basis we can represent a generally mixed quantum state with a density matrix: 
\begin{equation}\label{qubSt}
\hat{\rho}=\left[ \begin{array}{c c}
1-p & \gamma \\
\gamma^{*} & p \\
\end{array}\right],
\end{equation}
where $p\in[0,1]$ is the mean population of the excited state and $\gamma\in\mathbb{C}$ is a complex coherence term, with $\abs{\gamma}^{2}\leq p(1-p)$. The thermal attenuator channel $\Gamma_{\eta,\mathcal{N}}$ with $\eta \in [0,1]$ and $N\in[0,1/2]$, also known in the literature as generalized amplitude damping channel ~\cite{nChuangBOOK,amplitudeDamping}, acts on the matrix elements in the following way:
 \begin{align}
p  \xrightarrow{\Gamma_{\eta,\en}} & \; p'=  \eta p + (1-\eta) \en,   \label{GAD1} \\
\gamma  \xrightarrow{\Gamma_{\eta,\en}} &\; \gamma'=  \sqrt{\eta} \gamma,  \label{GAD2} 
\end{align}
and admits a representation according to the general structure given in Eqs. \eqref{phiDil} and \eqref{generalTA}. 
In this case the environment is given by a thermal two-level system: 
\begin{equation}\label{thQub}
\hat{\tau}=\left[\begin{array}{c c}
1-\en & 0 \\
0 & \en \\
\end{array}\right],
\end{equation}
where $\en \in [0,1/2]$ represents the dimensionless mean energy  and the bath hamiltonian is $\hat{H}_\mathrm{E}\propto\ketbra{1}$.  The unitary interaction is instead given by an energy-preserving rotation on the subspace $\{\ket{01},\ket{10}\}$ of the joint Hilbert space of the system: 
\begin{equation}
\hat{U}^{(\eta)}=\left[\begin{array}{cccc}
1&0&0&0\\
0&\sqrt{\eta}&\sqrt{1-\eta}&0\\
0&-\sqrt{1-\eta}&\sqrt{\eta}&0\\
0&0&0&1
\end{array}\right],
\end{equation}
physically inducing the hopping of excitations between the system and the environment. 
It is easy to check that, tracing out the environmental two-level system, we obtain the single qubit thermal attenuator channel defined in \eqref{GAD1} and \eqref{GAD2}.\\

A single mode of electromagnetic radiation instead is formally equivalent to an infinite-dimensional quantum harmonic oscillator. It can be described in terms of bosonic annihilation and creation operators $\hat a$ and $\hat a^\dag$, obeying the bosonic commutation relation $[\hat a, \hat a^\dag]=1$ or, equivalently, in terms of the quadratures $\hat q=(\hat a+\hat a^\dag)/\sqrt{2}$ and $\hat p=i (\hat a^\dag- \hat a)/\sqrt{2}$, such that $[\hat{q},\hat{p}]=i$. Given $n$ modes and introducing the vector of quadrature operators $\vecc{\hat{r}}=\left(\hat{q}_1, \hat{p}_1,..., \hat q_n , \hat p_n\right)^{\top}$,  one can define the characteristic function~\cite{RevGauss,serafiniBOOK} of a bosonic state $\hat{\rho}$:
\begin{equation}
\chi\left(\vecc{\xi}\right)=\tr{}{\hat{\rho}~e^{\vecc{\xi}^{T}\Omega^{T}\vecc{\hat{r}}}},
\end{equation}
where $\vecc{\xi}\in\mathbb{R}^{2n}$ and $\Omega=\left[\begin{array}{cc}0& 1 \\- 1&0 \end{array}\right]^{\oplus n}$ is the symplectic form. 
The most common bosonic states are Gaussian states, i.e. those whose characteristic function is Gaussian:
\begin{equation}
\chi_{\mathrm{G}}\left(\vecc{\xi}\right)=\exp\left\{-\frac{1}{4}\vecc{\xi}^{T}\Omega^{T}V\Omega\;\vecc{\xi}+\vecc{\xi}^{T}\Omega^{T}\vecc{m} \right\},
\end{equation}
uniquely determined by the first and second moments of the state $\hat{\rho}_{\mathrm{G}}$, given respectively by the vector of mean values $\vecc{m}$ and by the symmetric covariance matrix $V$ of the quadrature operators.
 The single-mode thermal attenuator channel $\mathcal E_{\eta,\en}$  has been extensively studied in the context of Gaussian quantum information~\cite{RevGauss,holevoBOOK,holGiovRev} 
 and its action on the first and second moments is given by:
 \begin{align}
\vecc{m}  \xrightarrow{\mathcal{E}{\eta,\en } }& \; \vecc{m}'=  \sqrt{\eta} \,\vecc{m},  \label{GTA1} \\
V  \xrightarrow{\mathcal{E}{\eta,\en }} &\; V'=  \eta V + (1-\eta) (2 \en + 1)\mathbf{1}_2, \label{GTA2} 
\end{align}
where $\eta\in[0,1]$ and $N\ge0$.
As for the qubit case, also this infinite-dimensional channel can be represented using the general picture of Eqs. \eqref{phiDil} and \eqref{generalTA}. Indeed, introducing the single-mode Gaussian 
thermal state $\hat{\tau}$ characterized by $\vecc{m}_{\hat{\tau}}=0$ and $V_{\hat{\tau}}=(2 \en+1)\mathbf{1}$, the thermal attenuator channel can be generated by a passive unitary interaction $\hat{U}$ that acts on the bosonic operators of the system, $\hat{a}$, and of the environment, $\hat{a}_{\mathrm{E}}$, as a beam splitter: 
 \begin{align}
U^\dag \hat a U &=  \sqrt{\eta} \hat{a} + \sqrt{1- \eta} \hat{a}_{\mathrm{E}}, \\
U^\dag \hat a_\mathrm{E} U &=  -\sqrt{1-\eta} \hat{a} + \sqrt{\eta}  \hat{a}_{\mathrm{E}}. 
\end{align}
It is easy to check that, tracing out the environmental mode, one recovers the definition of the single-mode thermal attenuator given in Eqs. \eqref{GTA1} and \eqref{GTA2}.

\subsection*{Known bounds for the Quantum Capacity of Thermal Attenuators}
The quantum capacity $Q(\Phi)$ of a channel $\Phi$ is defined as the maximum rate at which quantum information can be transferred by using  the channel  $N$ times with vanishing error in the limit $N\rightarrow\infty$. It is well known~\cite{lloydQuant,shorQuantum,qCap1,qCap2} that this quantity can be expressed as:
\begin{equation}\label{qCap}
Q(\Phi)=\lim_{N\rightarrow\infty}\frac{1}{N}\max_{\hat{\rho}\in\mathfrak{S}\left({\mathcal{H}_{\mathrm{A}}}^{\otimes N}\right)}J\left(\hat{\rho},\Phi^{\otimes N}\right),
\end{equation}
where
\begin{equation}\label{cohInfo}
J\left(\hat{\rho},\Phi\right)= S(\Phi(\hat \rho))-S(\tilde \Phi(\hat\rho))
\end{equation}
is the coherent information~\cite{cohInfo,holGiovRev} and $S(\hat{\rho})= -{\rm tr}\{ \hat{\rho} \log_{\mathrm{2}}  \hat{\rho}\} $ is the Von Neumann entropy; note that all logarithms henceforth are understood as base-2. Finally, $\tilde \Phi$ is the complementary channel~\cite{deg1}, obtained from the Stinespring dilation of $\Phi$ by tracing out the system instead of the environment, as detailed in the next subsection. %For example, the complementary channel to the thermal attenuator of \eqq{phiDil} is given by $\tilde{\Phi}_{\eta,N}(\rho)=\tr{B}{\hat{\rho}_{\mathrm{BF}}}$.
Because of the peculiar super-additivity phenomenon \cite{DiVincenzo1998,Smith2007,zeroQCap,hastings2009,supAddQCap}, the so called single-letter capacity  
\begin{align}\label{q1}
Q_{1}(\Phi)=\max_{\hat{\rho}\in\mathcal{H}_{\mathrm{A}}}J(\hat{\rho},\Phi)
\end{align}
is in general smaller than the actual capacity of the channel.  

This fact directly gives a lower bound for the quantum capacity of the qubit thermal attenuator: 
\begin{equation}\label{qubLowBound}
Q(\Gamma_{\eta,N}) \ge  Q_{1}(\Gamma_{\eta,N})=\max_{p} J \left ( \hat{\rho}=\left[ \begin{array}{c c}
1-p & 0 \\
0& p \\
\end{array}\right] ,\Gamma_{\eta,N} \right)
\end{equation} 
where the maximization is only with respect to  $p\in [0,1]$, since one can check numerically that for this channel optimal states are diagonal; this quantity can be easily numerically computed for all values of $\eta$ and $N$.
Similarly, a lower bound can be obtained also for the bosonic counterpart of the thermal attenuator by restricting the optimization over the class of Gaussian input states \cite{holWer}:
\begin{align}\label{GaussLowBound}
Q(\mathcal E_{\eta,N}) \ge  Q_{1}(\mathcal E_{\eta,N})  &\ge \max_{\hat \rho_\mathrm{G}} J (\hat \rho_\mathrm{G} ,\mathcal E_{\eta,N})  \nonumber \\
& =\max \left\{0, \log_{\mathrm{2}} \frac{\eta}{1-\eta}-g(\en) \right\},
\end{align} 
where $\rho_\mathrm{G}$ varies over the set Gaussian states and $g(\en)=(\en+1)\log_{\mathrm{2}} (\en+1)-\en\log_{\mathrm{2}} \en$ corresponds to the entropy of the thermal state of the environment.

Whether the right-hand-sides of \eqref{qubLowBound} and \eqref{GaussLowBound} are equal or not to the true quantum capacity of the associated channels is still an important open problem in quantum information. It can be shown \cite{deg1,amplitudeDamping,qCapLossy}, that  
for a zero-temperature environment this is the case, i.e. for $N=0$ all inequalities in \eqref{qubLowBound} and \eqref{GaussLowBound} are saturated, giving the quantum capacity of both the qubit and the Gaussian attenuators.  
For $N>0$ instead, the capacity is still unknown, apart from some upper bounds. For the qubit case we are not aware of any upper bounds proposed in the literature, while for the Gaussian thermal attenuator the best bounds at the moment are those recently introduced in \cite{plob,wildeDecDirettaBound}, which can be combined to get the following expression:
{\color{red}\begin{eqnarray}\label{boundPLOB}
&&Q(\mathcal E_{\eta,N}) \le \min \left\{Q_{\mathrm{PLOB}}(\mathcal E_{\eta,N}), Q_{\mathrm{SWAT}}(\mathcal E_{\eta,N})\right\},\\
&&Q_{\mathrm{PLOB}}(\mathcal E_{\eta,N})=\max\{0,- \log_{\mathrm{2}} [(1-\eta)\eta^N]-g(\en)\},\nonumber\\
&&Q_{\mathrm{SWAT}}(\mathcal E_{\eta,N})=\max\{0,\log_{\mathrm{2}} [\eta/(1-\eta)]-\log_{\mathrm{2}} [N+1]\}.\nonumber
\end{eqnarray}
Let us note that $Q_{\mathrm{PLOB}}$ is actually a bound on the quantum capacity assisted by two-way classical communication~\cite{plob} and thus trivially bounds also the simpler unassisted capacity discussed in this article (strong-converse bounds for the two-way capacity where derived, e.g., in ~\cite{wilde2017,Christandl2017}). The other bound instead, $Q_{\mathrm{SWAT}}$, is itself a bound on the unassisted capacity and it has been shown~\cite{wildeDecDirettaBound} to beat other possible bounds based on $\epsilon$-degradability~\cite{Sutter2017}.}
In the next sections we are going to derive new upper bounds which are significantly closer to the lower limits \eqref{qubLowBound} and \eqref{GaussLowBound}, especially in the low temperature regime.

\subsection*{The extended channel}\label{sec:extCh}
In this subsection we first review the notions of degradability and weak degradability and then we introduce an extended version of thermal attenuator maps, whose quantum capacity is easier to compute and can be used as a useful upper bound. 

Let us come back to the description of a generic attenuator map $\Phi_{\eta,\en}$,  valid both for qubit and bosonic systems, that we have previously defined in Eqs.~(\ref{phiDil}, \ref{generalTA}).
If, instead of tracing out the environment as done in Eq. \eqref{generalTA}, we trace out the system, we get what has been defined in \cite{weakDeg,antiDeg} as the weakly-complementary channel:
\begin{equation} \label{wComp}
\bar \Phi_{\eta,\en}\left(\hat{\rho}_{\mathrm{A}}\right)=\tr{B}{\hat{\rho}_{\mathrm{BF}}},
\end{equation}
physically representing the flow of information from the system into the environment. Notice that this is different form the standard notion of complementary channel:
\begin{equation}
\tilde \Phi_{\eta,\en}\left(\hat{\rho}_{\mathrm{A}}\right)=\tr{B}{\hat{\rho}_{\mathrm{BFE'}}},
\end{equation}
where 
\begin{equation}\label{phiExtDil}
\hat{\rho}_{\mathrm{BFE'}}=\hat{U}^{(\eta)}_{\mathrm{AE}\rightarrow \mathrm{BF}}\left(\hat{\rho}_{\mathrm{A}}\otimes\ketbra{\tau}_{\mathrm{EE'}}\right)\hat{U}_{\mathrm{AE}\rightarrow \mathrm{BF}}^{(\eta) \dag}
\end{equation}
and $| \tau \rangle_{\mathrm{EE'}}$ is a purification of the environment, i.e.  $\hat{\tau}_{\mathrm{E}}=\tr{E'}{ \ketbra{\tau}_{\mathrm{EE'}} }$.
The weak and standard complementary channels become equivalent (up to a trivial isometry) only in the particular case in which the environment is initially pure (zero-temperature limit).
Finally, we also remark that the two different types of complementarity induce different definitions of degradability. A generic channel $\Phi$ is degradable \cite{deg1,deg2},  if there exists another quantum channel $\Delta$ such that $\tilde \Phi= \Delta \circ  \Phi$. Similarly, a generic channel $\Phi$ is weakly degradable \cite{weakDeg,antiDeg}, if there exists another quantum channel $\Delta$ such that $\bar \Phi= \Delta \circ  \Phi$. For degradable channels, the capacity is additive and much easier to determine. Unfortunately, typical models of quantum attenuators, as the qubit and the bosonic examples considered in this work, are degradable only for $N=0$ but become only weakly degradable for $N>0$ and this is the main reason behind the hardness in computing their quantum capacity.  

In order to circumvent this problem, we define the extended version of a thermal attenuator channel as
\begin{equation} \label{extended}
\Phi_{\eta,\en}^{\mathrm{e}}\left(\hat{\rho}_{\mathrm{A}}\right)=\tr{F}{\hat{\rho}_{\mathrm{BFE'}}},
\end{equation}
where $\hat{\rho}_{\mathrm{BFE'}}$ is the global state defined in \eqref{phiExtDil}. In other words, $\Phi_{\eta,\en}^{\mathrm{e}}$ represents a situation in which Bob has access not only to the output system B but also to the purifying part E' of the environment, see Fig.~\ref{fg1}. A remarkable fact is that, locally,  the auxiliary system E' remains always in the initial thermal state because it is unaffected by the dynamics; however E' can be correlated with B and this fact can be exploited by Bob to retrieve more quantum information. In general, since trowing away E' can only reduce the quantum capacity, one always has $Q(\Phi_{\eta,\en}) \le Q(\Phi_{\eta,\en}^{\mathrm{e}})$.\\
The advantage of dealing with the extended channel $\Phi_{\eta,\en}^{\mathrm{e}}$ is that it is degradable whenever the original channel $\Phi_{\eta,\en}$ is weakly degradable. This fact follows straightforwardly from the observation that the complementary channel of $\Phi_{\eta,\en}^{\mathrm{e}}$ is the weakly-complementary channel of $\Phi_{\eta,\en}$, i.e.: 
\begin{equation}\label{property1}
\tilde{\Phi}_{\eta,\en}^{e}\left(\hat{\rho}_A\right)= \tr{BE'}{\hat{\rho}_{\mathrm{BFE'}}}= \bar{\Phi}_{\eta,\en}\left(\hat{\rho}_A\right).
\end{equation}
The degradability of $\Phi_{\eta,\en}^{\mathrm{e}}$ significantly simplifies the evaluation of $Q(\Phi_{\eta,\en}^{\mathrm{e}})$ and provides a very useful upper bound for the quantum capacity of thermal attenuators:
\begin{equation}\label{generalBound}
Q(\Phi_{\eta,\en}) \le Q(\Phi_{\eta,\en}^{\mathrm{e}})=Q_1(\Phi_{\eta,\en}^{\mathrm{e}}),
\end{equation}
where in the last step we used the additivity property valid for all degradable channels \cite{deg1,deg2}.\\
Another useful property of the extended channel, which follows from Eq.\ \eqref{property1} and the definition of coherent information, \eqq{cohInfo}, is the following:
\begin{equation}\label{property2}
J\left(\rho, \Phi_{\eta,\en}^{\mathrm{e}}\right) = - J\left(\rho, \bar{\Phi}_{\eta,\en}\right).
\end{equation}
relating the coherent information of the extended and of the weakly-complementary channels.\\
Below we compute more explicitly the previous bound \eqref{generalBound} for the specific cases of discrete- and continuous-variable thermal attenuators.\\

\begin{figure}[t!]
\includegraphics[scale=.55,trim={3cm 4cm 1.5cm 2cm},clip]{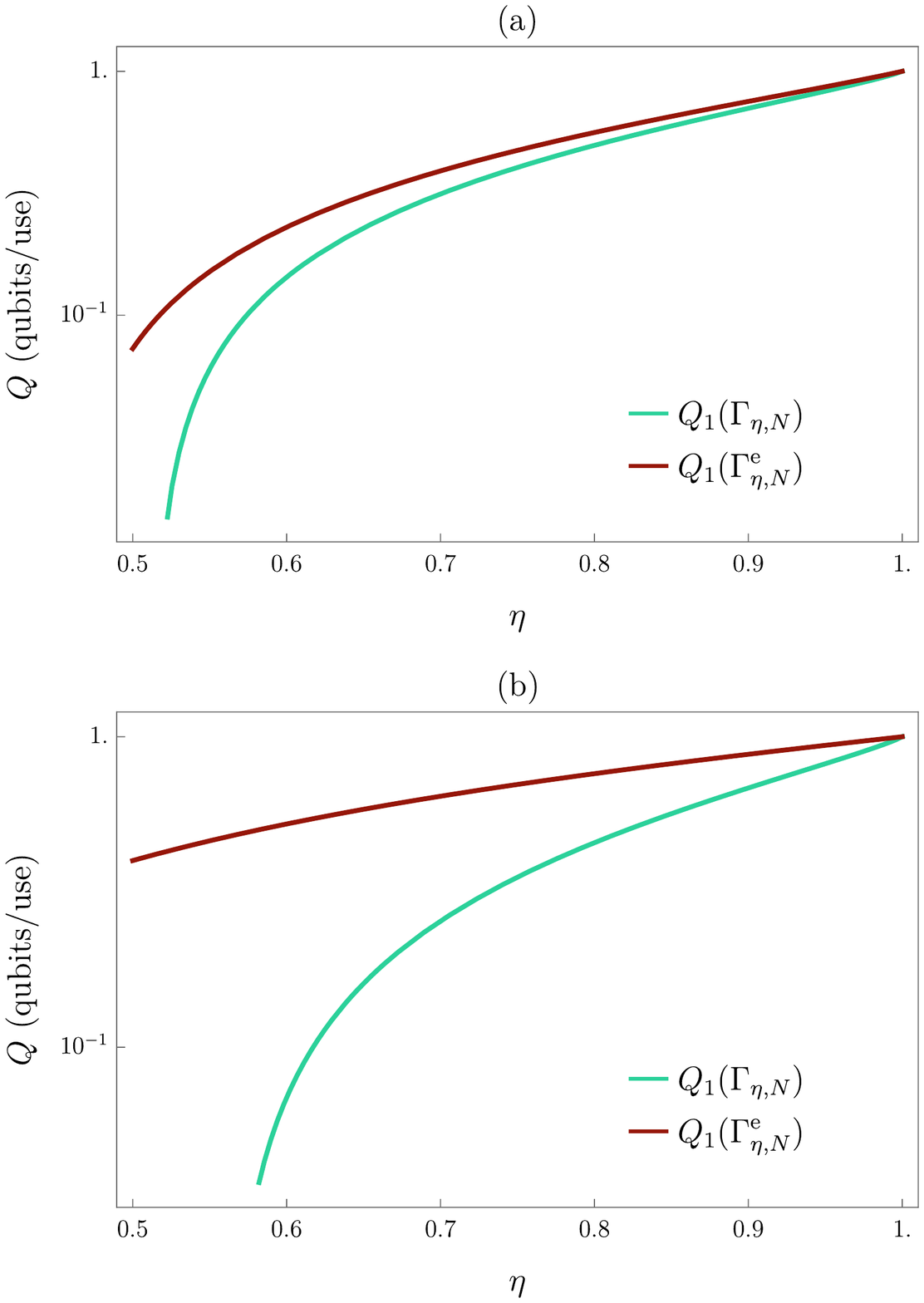}
{\color{red}\caption{Bounds of the quantum capacity of a qubit thermal attenuator. \\ Plot (log-linear scale) of the lower bound $Q_1(\Gamma_{\eta,N})$ (green line), \eqq{qubLowBound}, and upper bound $Q_1(\Gamma_{\eta,N}^\mathrm{e})$ (red line), \eqq{qubUpBound}, of the quantum capacity of the qubit thermal attenuator channel as a function of the attenuation parameter $\eta$ for two values of noise: (a) $N=0.01$, (b) $N=0.1$. The lower bound is given by the single-letter capacity of the channel, whereas the upper bound by the capacity of the extended channel, which equals its single-letter expression because of degradability. The two bounds are close for small noise and weak attenuation.}\label{fig:plotQub}}
\end{figure}

In the case of two-level systems, the purification of the thermal state given in \eqq{thQub} is
\begin{equation}
\ket{\tau}=\sqrt{1-\en}\ket{00}+ \sqrt{\en}\ket{11}.
\end{equation}
The channel $\Gamma_{\eta,\en}$ can be weakly degraded to $\bar \Gamma_{\eta,\en}$ by the composite map $\Delta=\Psi_{\mu}\circ\Gamma_{\eta',\en}$, where $\eta'=(1-\eta)/\eta$. Here $\Psi_{\mu}$ is a phase-damping channel~\cite{nChuangBOOK} of parameter $\mu=1-2\en$, which acts on the generic qubit state of \eqq{qubSt} by damping the coherence matrix element as $\gamma\mapsto\mu\gamma$ while leaving the population $p$ constant. Hence the extended channel $\Gamma_{\eta,\en}^{\mathrm{e}}$ defined as in \eqref{extended} is degradable and from \eqref{generalBound} we get 
\begin{equation}\label{qubUpBound}
 Q(\Gamma_{\eta,N}) \le Q_1(\Gamma_{\eta,N}^\mathrm{e}) = \max_{p} J \left ( \hat{\rho}=\left[ \begin{array}{c c}
1-p & 0 \\
0& p \\
\end{array}\right] ,\Gamma^\mathrm{e}_{\eta,N} \right),
\end{equation} 
where the optimization over the single parameter $p$ can be efficiently performed numerically for all values of $\eta$ and $N$, giving the result presented in Fig.~\ref{fig:plotQub}. In this case the fact that we can reduce the optimization over diagonal input states follows from the symmetry of the coherent information under the matrix-element flipping $\gamma \rightarrow - \gamma$ and from the concavity of the coherent information for degradable channels \cite{Yard2005}.\\
We note that, by construction, the gap between the lower  \eqref{qubLowBound}  and the upper \eqref{qubUpBound}  bounds closes in the limit $\en \rightarrow 0$, where we recover the zero-temperature capacity of the amplitude damping channel consistently with \cite{amplitudeDamping}.
\cite{holWer}.
\\

In the case of bosonic systems instead, the purification of the thermal environmental mode $\hat\tau$ with first moments $\vecc{m}_{\hat{\tau}}=0$ and covariance matrix $V_{\hat{\tau}}=(2 \en+1)\mathbf{1}$ is a Gaussian two-mode squeezed state $\ket{\tau}$  ~\cite{RevGauss}, characterized by $\vecc{m}_{\ket{\tau}}=0$ and
\begin{equation}
V_{\ket{\tau}}=\left(\begin{array}{cc} V_{\hat\tau}&\sigma_{3} \sqrt{ V_{\hat\tau}^2-\mathbf{1}_2} \\ \sigma_{3}\sqrt{V_{\hat\tau}^2-\mathbf{1}_2}&V_{\hat\tau}\end{array}\right),
\end{equation}
where $\sigma_{3}=\operatorname{diag}(1,-1)$ is the third Pauli matrix. 
The thermal attenuator $\mathcal E_{\eta,\en}$ is weakly degradable \cite{weakDeg,antiDeg}, since its weakly complementary channel can be expressed as $\bar{\mathcal E}_{\eta,N}=\Delta\circ \mathcal E_{\eta,N}$ where $\Delta=\mathcal E_{\eta',\en}$, with $\eta'=(1-\eta)/\eta$. 
Hence $\mathcal E_{\eta,\en}^\mathrm{e}$ is degradable and we can apply the general upper bound \eqref{generalBound}.  Moreover, as shown in~\cite{wolf2006,qCapLossy}, the quantum capacity of a degradable Gaussian channel is maximized by Gaussian states with fixed second moments, so that we can write 
\begin{align}\label{upBos0}
Q\left(\mathcal E_{\eta,\en}\right) \le Q_1\left(\mathcal E_{\eta,\en}^\mathrm{e}\right)&= \max_{\hat{\rho}_\mathrm{G}} J (\hat{\rho}_\mathrm{G} ,\mathcal E_{\eta,N}^\mathrm{e}) % 
\end{align} 
reducing the problem to a tractable Gaussian optimization.
Since the coherent information of the extended channel is concave and symmetric with respect to phase-space rotations and translations, for a fixed energy $n$ it is maximized by the thermal state $\hat{\tau}_n$ with covariance matrix $V_{\hat{\tau}_n}=(1+2n)\mathbf{1}_2$, see Ref.~\cite{wildeThermOpt}. Therefore,
without energy constraint we have
\begin{align}\label{upBos0}
Q\left(\mathcal E_{\eta,\en}\right) \le Q_1\left(\mathcal E_{\eta,\en}^\mathrm{e}\right)&=\lim_{n\rightarrow \infty} J (\hat{\tau}_n ,\mathcal E_{\eta,N}^\mathrm{e}).
\end{align}
The last term can be explicitly computed by using the standard formalism of Gaussian states; more simply, we can relate this quantity to the results of 
Indeed, from the property given in Eq.\ \eqref{property2}, we have
\begin{equation}
J(\hat{\tau}_n ,\mathcal E_{\eta,N}^\mathrm{e})=-J(\hat{\tau}_n ,\bar{\mathcal E}_{\eta,N})=-J(\hat{\tau}_n ,\mathcal E_{1-\eta,N}),
\end{equation}
but the last term is simply the negative of the coherent information of a thermal attenuator of transmissivity $\eta'=1-\eta$ and a thermal input state, a quantity which has been already computed in \cite{holWer}. In the limit of $n \rightarrow \infty$, we get our desired upper bound:
\begin{align}\label{upBos}
Q\left(\mathcal E_{\eta,\en}\right) \le Q_1\left(\mathcal E_{\eta,\en}^\mathrm{e}\right)&= \max \left\{0, \log_{\mathrm{2}} \frac{\eta}{1-\eta}+g(\en) \right\},
\end{align}
where $g(\en)=(\en+1)\log_{\mathrm{2}} (\en+1)-\en\log_{\mathrm{2}} \en$, shown in Fig.~\ref{fig:plotBos}.
Comparing the upper bound \eqref{upBos} with the lower bound \eqref{GaussLowBound} we observe that we can determine the quantum capacity of the thermal attenuator up to an uncertainty of $2g(\en)$, which vanishes in the limit of small thermal noise. For the special case $\en=0$, the gap closes and we recover the capacity
of the pure lossy channel consistently with the previous results of Ref.\ \cite{qCapLossy}.

\subsection*{Twisted decomposition of Gaussian attenuators}\label{sec:altBound}

In this section, through a completely different method, we derive a bound for the quantum capacity which is tighter than $Q\left(\Phi_{\eta,\en}^{\mathrm{e}}\right)$ but applies only to the bosonic version of thermal attenuators. 

Let us first introduce a second kind of thermal Gaussian channel: the single-mode amplifier $\mathcal{A}_{\kappa,\en}$ \cite{RevGauss}, which combines the input state and the usual thermal state $\hat{\tau}$ of energy $\en$ through a two-mode squeezing interaction with gain $\kappa>1$. Tracing out the environment, the first and second moments of the quantum state transform in the following way:
 \begin{align}\label{GAD}
\vecc{m}  \xrightarrow{\mathcal{A}_{\kappa,\en}} & \; \vecc{m}'=  \sqrt{\kappa} \,\vecc{m},  \\
V  \xrightarrow{\mathcal{A}_{\kappa,\en}} &\; V'=  \kappa V + (\kappa-1) (2\en + 1)\mathbf{1}.  
\end{align}
In the particular case in which the environment is at zero temperature, i.e. for $\en=0$, the channel $\mathcal A_{\kappa,0}$ is called quantum-limited amplifier. 
\begin{figure}[t!]
\includegraphics[scale=.48,trim={2cm 1.5cm 2cm 2cm},clip]{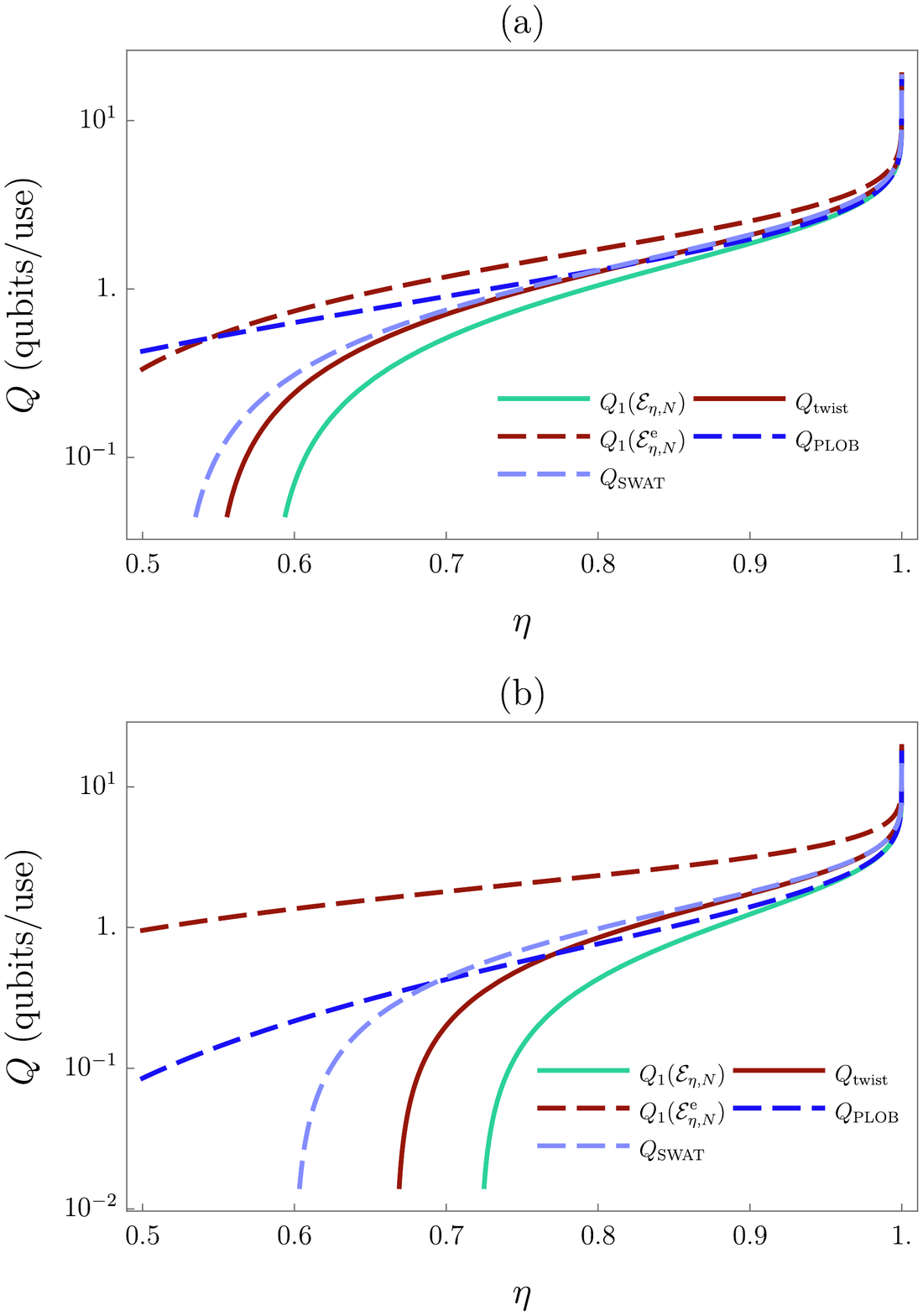}
{\color{red}\caption{Bounds of the quantum capacity of a bosonic thermal attenuator. \\ Plot (log-linear scale) of the lower bound $Q_{1}(\mathcal{E}_{\eta,N})$ (green line), \eqq{GaussLowBound}, and several upper bounds of the quantum capacity of the bosonic Gaussian thermal attenuator channel as a function of the attenuation parameter $\tau$ for two values of noise: (a) $N=0.1$, (b) $N=0.5$. The upper bounds are given by: the capacity of the extended channel $Q_{1}(\mathcal{E}_{\eta,N}^{\mathrm{e}})$ (red dashed line), \eqq{upBos}; the capacity of the attenuator channel of the twisted decomposition $Q_{\mathrm{twist}}$ (red line), \eqq{twistBound}; the upper bounds $Q_{\mathrm{PLOB}}$ of \eqq{boundPLOB} derived from Ref.~\cite{plob} (blue dashed line) and $Q_{\mathrm{SWAT}}$ of Ref.~\cite{wildeDecDirettaBound} (light-blue dashed line). Note that the best upper bound at small noise values is provided by our $Q_{\mathrm{twist}}$. As the noise increases, $Q_{\mathrm{PLOB}}$ starts beating the former for weak attenuation, while $Q_{\mathrm{SWAT}}$ remains always strictly larger than our bound.}\label{fig:plotBos}}
\end{figure}

It can be shown that all phase-insensitive Gaussian channels can be decomposed as a quantum-limited attenuator followed by a quantum-limited amplifier \cite{deco2,weakDeg}, with important implications for their classical capacity \cite{gaussOpt,gaussCap,gaussMaj}. 
In this work we introduce a twisted version of this decomposition in which the order of the attenuator and of the amplifier is inverted, which is quite useful for bounding the quantum capacity of thermal attenuators.  

\theoremstyle{remark}
\newtheorem{lemma}{Lemma}
\begin{lemma}
Every thermal attenuator $\mathcal E_{\eta,\en}$ that is not entanglement-breaking can be decomposed as a quantum-limited amplifier followed by a quantum-limited attenuator:
\begin{equation}\label{twist}
\mathcal E_{\eta,\en}=  \mathcal E_{\eta',0}  \circ \mathcal A_{\kappa',0} ,
\end{equation} 
with attenuation and gain coefficients given by
\begin{equation}\label{coeff}
\eta'=\eta - \en (1-\eta) , \quad \kappa'= \eta/\eta'
\end{equation} 
\end{lemma}
The proof can be obtained by direct substitution and using the fact that non-entanglement-breaking attenuators are characterized by the condition $\en < \eta/(1-\eta)$ \cite{holGiovRev,holevo2008} and so both coefficients in \eqref{coeff} are positive and well defined. As shown in the Methods Section, the previous decomposition can be generalized to all phase-insensitive Gaussian channels including thermal amplifiers and additive Gaussian noise channels. It is important to remark that, differently from the decomposition introduced in Ref. \cite{deco2} and employed in Ref.~\cite{wildeDecDirettaBound}, our twisted version does not apply to entanglement-breaking channels. For the purposes of this work, this is not a restriction since all entanglement-breaking channels trivially have zero quantum capacity and we can exclude them from our analysis. 

\begin{figure}[t!]
\includegraphics[scale=1,trim={6.5cm 17.5cm 6.5cm 1.5cm},clip]{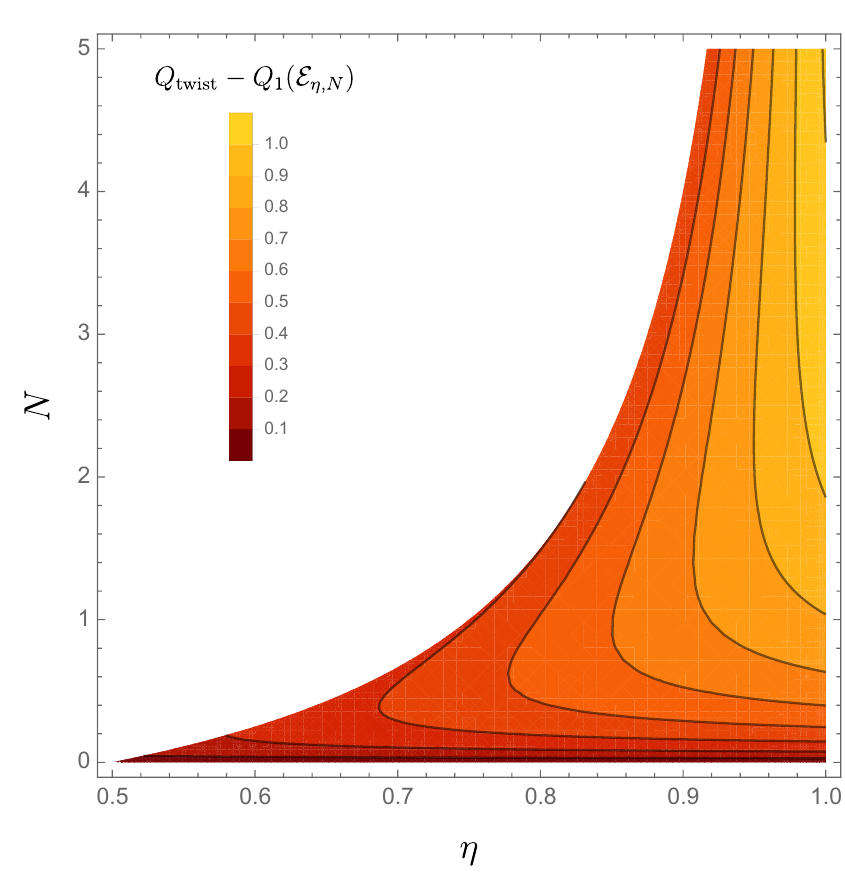}
{\color{red} \caption{Best-approximation accuracy of the quantum capacity of a bosonic thermal attenuator. \\ Contour plot of the difference between $Q_\mathrm{twist}$ of \eqq{twistBound}, i.e., the twisted-decomposition upper bound of the quantum capacity of the bosonic Gaussian thermal attenuator, and the lower bound $Q_{1}(\mathcal{E}_{\eta,N})$ of \eqq{GaussLowBound}, i.e., the single-letter capacity of the channel, as a function of the attenuation parameter $\eta\in[0.5,1]$ and noise values $N\in[0,5]$. The white region corresponds to zero capacity. Observe that the approximation is tight in the small-noise region and, at higher values of noise, in the strong-attenuation region. Note that, as shown in Fig.~\ref{fig:plotBos}b, in the opposite regime of high values of $N$ and $\eta$, the quantum capacity is better upper bounded by \eqref{boundPLOB}.\label{fig:contBos}}
}
\end{figure}

Now, given a thermal attenuator with $\en < \eta/(1-\eta)$, we make use of the twisted decomposition \eqref{twist} obtaining
\begin{align} \label{bottle}
Q(\Phi_{\eta,\en})= Q( \Phi_{\eta',0}  \circ \mathcal A_{\kappa',0} ) &\le Q( \Phi_{\eta',0}) \\
                                                                       &=\max \left\{0,\log_{\mathrm{2}}   \frac{\eta'}{1-\eta'}\right\}, 
\end{align}
where we used the ``bottleneck" inequality $Q(\Phi_1 \circ \Phi_2 )\le \min\{ Q(\Phi_1), Q(\Phi_2)  \}$ and the exact expression for the capacity of the quantum-limited attenuator \cite{qCapLossy}.
Substituting the value of $\eta'$ of Eq. \eqref{coeff} into \eqref{bottle}, we get our desired upper bound
\begin{align} \label{twistBound}
Q(\Phi_{\eta,\en}) \le Q_{\mathrm{twist}}=\max \left\{0, \log_{\mathrm{2}}  \frac{\eta-\en (1-\eta)}{ (1+\en) (1-\eta)} \right\}.
\end{align}
One can easily check that this last bound is always better than the one derived in Eq.\ \eqref{upBos} and the bound $Q_{\mathrm{SWAT}}$ of Eq.~\eqref{boundPLOB}. Moreover, for sufficiently small $\eta$ or for sufficiently small $\en$, it outperforms also the bound $Q_{\mathrm{PLOB}}$ of Eq.~\eqref{boundPLOB}, see Fig.~\ref{fig:plotBos}. {\color{red} By combining our result, $Q_{\mathrm{twist}}$, with $Q_{\mathrm{PLOB}}$ and with the lower bound $Q_{1}(\Phi_{\eta,\en})$, the quantum capacity is now constrained within a very small uncertainty window.}
Figure~\ref{fig:contBos} shows the tiny gap existing between our new upper bound \eqref{twistBound} based on the twisted decomposition and the lower bound \eqref{GaussLowBound}.

\section*{Discussion}\label{sec:conc}
In this article we computed some upper bounds on the quantum capacity of thermal attenuator channels, making use of an extended channel whose degradability properties are preserved when the environment has non-zero mean energy. This method gave interesting bounds in both the qubit and bosonic case, which are tight in the low temperature limit. Our method is quite general since it can be applied to any weakly-degradable channel that admits a physical dilation with a mixed environmental state, not necessarily thermal. For example one can apply this method straightforwardly to the bosonic thermal amplifier, though in this case the previously known upper bound \cite{plob} is very tight and cannot be improved in this way. The second method we employed is less general, since it relies on a specific decomposition of thermal attenuators, but provides a better upper bound. Moreover, the twisted decomposition of Gaussian channels that we introduced in this work is an interesting result in itself which could find application in other contexts.

Our methods are of general interest for computing also other information capacities. Indeed, the channels that we introduce, i.e., the extended channel and those constituting the twisted decomposition, are by construction less noisy than the TA, in the sense that the latter can be obtained by any of the former via concatenation with another channel. This key property allows in principle to upper-bound any information capacity of a TA with that of any of the channels we introduced. Of course, this may turn out to be very difficult in practice, depending on the kind of capacity that we are interested in. For example, a bound on the private capacity seems straightforward to derive, whereas it would require more efforts to bound the two-way and the strong-converse quantum capacities, the former because of the lack of a closed expression and the latter because of the difficult regularization involved in its formula. For these reasons, we believe that the channels we introduced are worth investigating also in the context of bounding other information capacities and may provide further interesting results.

Combining the results of this work with other previously known bounds, we can now estimate the value of the quantum capacity of thermal attenuator channels up to corrections which are irrelevant for most practical purposes. 

\cleardoublepage
\section*{Methods}

\subsection*{Computation of the upper bound for Qubit TA}
The capacity of the extended qubit attenuator $\Gamma_{\eta,N}^\mathrm{e}$ can be computed by maximizing the coherent information of the channel, which is additive, as discussed in the main text:
\begin{align}
Q_{1}(\Gamma_{\eta,N}^\mathrm{e})&=\max_{\hat{\rho}}\left[S\left(\Gamma_{\eta,N}^\mathrm{e}\left(\hat{\rho}\right)\right)-S\left(\tilde{\Gamma}_{\eta,N}^\mathrm{e}\left(\hat{\rho}\right)\right)\right] \nonumber \\
&=\max_{p,\gamma}J_{\eta,N}(p,\gamma)
\end{align}
and we have defined for simplicity of notation 
\begin{equation}
J_{\eta,N}(p,\gamma)=J \left ( \hat{\rho}=\left[ \begin{array}{c c}
1-p & \gamma \\
\gamma^*& p \\
\end{array}\right] ,\Gamma^\mathrm{e}_{\eta,N} \right),
\end{equation} 
depending on the parameters $p,\gamma$ of the input qubit and on the channel parameters $\eta,N$.
Recall that the extended channel $\Gamma_{\eta,N}^\mathrm{e}$ maps states of the system A to states of the joint system BE' that includes the purifying part of the environment. Conversely, its complementary $\tilde{\Gamma}_{\eta,N}^\mathrm{e}$ maps states of A to states of the interacting part of the environment F. Therefore to compute their entropies we start by writing the joint state of the system AEE', in the basis $\{\ket{0},\ket{1}\}^{\otimes 3}$:
\begin{widetext}
\begin{equation}\small
\hat{\rho}_\mathrm{AEE'}=\left(
\begin{array}{cccccccc}
 (1-N) (1-p) & 0 & 0 & \sqrt{(1-N) N} (1-p) & \gamma (1-N) & 0 & 0 & \sqrt{(1-N) N} \gamma  \\
 0 & 0 & 0 & 0 & 0 & 0 & 0 & 0 \\
 0 & 0 & 0 & 0 & 0 & 0 & 0 & 0 \\
 \sqrt{(1-N) N} (1-p) & 0 & 0 & N(1-p) & \sqrt{(1-N) N} \gamma  & 0 & 0 & N \gamma  \\
 (1-N) \gamma ^* & 0 & 0 & \sqrt{(1-N) N} \gamma ^* & p-N p & 0 & 0 & \sqrt{(1-N) N} p \\
 0 & 0 & 0 & 0 & 0 & 0 & 0 & 0 \\
 0 & 0 & 0 & 0 & 0 & 0 & 0 & 0 \\
 \sqrt{(1-N) N} \gamma ^* & 0 & 0 & N \gamma ^* & \sqrt{(1-N) N} p & 0 & 0 & N p \\
\end{array}
\right).
\end{equation}
Next, we compute its evolved $\hat{\rho}_{\mathrm{BFE'}}$ under the action of $\mathcal{U}^{(\eta)}_{\mathrm{AE}\rightarrow \mathrm{BF}}\otimes\mathbf{I}_\mathrm{E'}$ and the marginals with respect to the bipartition BE'-F, which correspond to the output states of the two channels:
\begin{align}
&\small\Gamma_{\eta,N}^\mathrm{e}\left(\hat{\rho}\right)=\left(
\begin{array}{cccc}
 (1-N) (1-p \eta ) & \sqrt{(1-N) N (1-\eta ) \eta } \gamma ^* & \gamma  (1-N)\sqrt{\eta } & (2 p-1) \sqrt{(1-N) N (1-\eta )} \\
 \gamma  \sqrt{(1-N) N (1-\eta ) \eta } & N (1-p)\eta  & 0 & N \gamma  \sqrt{\eta } \\
 (1-N) \sqrt{\eta } \gamma ^* & 0 & p (1-N) \eta  & -\sqrt{(1-N) N (1-\eta ) \eta } \gamma ^* \\
 (2 p-1) \sqrt{(1-N) N (1-\eta )} & N \sqrt{\eta } \gamma ^* & -\gamma  \sqrt{(1-N) N (1-\eta ) \eta } & (p-1) \eta  N+N \\
\end{array}
\right),\nonumber\\
\\
&\small\tilde{\Gamma}_{\eta,N}^\mathrm{e}\left(\hat{\rho}\right)=\left(
\begin{array}{cc}
 -p (1-\eta )-N \eta +1 & (1-2 N) \gamma  \sqrt{1-\eta } \\
 (1-2 N) \sqrt{1-\eta } \gamma ^* & -\eta  p+p+N \eta  \\
\end{array}
\right).\nonumber
\end{align}
\end{widetext}
The entropies of these states can be computed numerically and it can be checked that they are invariant under phase-flip, i.e., $\gamma\rightarrow-\gamma$. Hence we obtain an expression of the coherent information that is an even function of $\gamma$ and write, following \cite{wildeBOOK}:
\begin{equation}
J_{\eta,N}(p,\gamma)=\frac{J_{\eta,N}(p,\gamma)+J_{\eta,N}(p,-\gamma)}{2}\leq J_{\eta,N}(p,0),
\end{equation}
where the inequality follows from the concavity of the mutual information as a function of the input state, which holds since the channel is degradable~\cite{Yard2005}. Hence we have restricted the optimization to diagonal states in the chosen basis, i.e., on the single parameter $p$ for fixed $\eta,N$:
\begin{equation}
Q_{1}(\Gamma_{\eta,N}^\mathrm{e})=\max_{p}J_{\eta,N}(p,0),
\end{equation}
for $\eta>1/2$ and $0$ otherwise. The latter expression can be easily solved numerically, as shown by the plots in the main text.

\subsection*{Twisted decomposition of phase-insensitive Gaussian channels}
Here we generalize the twisted decomposition introduced in the main text for bosonic thermal attenuators to the more general class of phase-insensitive Gaussian channels $\mathcal G_{\tau,y}$, defined by the following action on the first and second moments of single-mode Gaussian states \cite{RevGauss}:
 \begin{align}
\vecc{m}  \xrightarrow{\mathcal{G}{\tau,y } }& \; \vecc{m}'=  \sqrt{\tau} \,\vecc{m},  \label{G1} \\
V  \xrightarrow{\mathcal{G}{\tau,y }} &\; V'=  \tau V + y \mathbf{1}_2, \label{G2} 
\end{align}
where $\tau\ge 0$ is a generalized transmissivity and $y\ge |1-\tau| $ is a noise parameter \cite{holGiovRev}. This family includes the thermal attenuator for $0\le\tau<1$, the thermal amplifier for $\tau>1$, and the additive Gaussian noise channel for $\tau=1$. If $y=|1-\tau|$, the channel introduces the minimum noise allowed by quantum mechanics and is said to be quantum-limited.
On the other hand, it can be shown \cite{holGiovRev,holevo2008} that a phase-insensitive Gaussian channel is entanglement-breaking if and only if $y\ge 1+\tau$,
which determines a noise threshold above which the channel has trivially zero quantum capacity.
Below the entanglement-breaking threshold, the following decomposition holds.
 
\begin{lemma}
Every phase-insensitive Gaussian channel $\mathcal G_{\tau,y}$ which is not entanglement-breaking ($y< 1+\tau$), can be decomposed as a quantum-limited amplifier followed by a quantum-limited attenuator:
\begin{equation}\label{twist2}
\mathcal G_{\tau,y}=  \mathcal G_{\eta',1-\eta'} \circ \mathcal G_{\kappa',\kappa'-1} =\mathcal E_{\eta',0}  \circ \mathcal A_{\kappa',0} ,
\end{equation} 
with attenuation and gain coefficients given by
\begin{equation}\label{coeff2}
\eta'=(1+\tau - y)/2 , \quad \kappa'= \tau/\eta'.
\end{equation} 
\end{lemma}
The proof follows by direct substitution of the parameters \eqref{coeff2} into \eqref{twist2} and from the application of Eqs.~(\ref{G1}, \ref{G2}). 
Moreover, the hypothesis $y< 1+\tau$ is necessary since it ensures the positivity of both the attenuation and the gain parameters $\eta'$ and $\kappa '$.

\section*{Data availability}
No datasets were generated or analysed during the current study.

%\bibliographystyle{naturemag_noURL}
%\bibliography{library}

\begin{thebibliography}{10}
\section*{References}
\expandafter\ifx\csname url\endcsname\relax
  \def\url#1{\texttt{#1}}\fi
\expandafter\ifx\csname urlprefix\endcsname\relax\def\urlprefix{URL }\fi
\providecommand{\bibinfo}[2]{#2}
\providecommand{\eprint}[2][]{\url{#2}}

\bibitem{shannonSeminal}
\bibinfo{author}{Shannon, C.~E.}
\newblock \bibinfo{title}{{A mathematical theory of communication}}.
\newblock \emph{\bibinfo{journal}{Bell Syst. Tech. J.}}
  \textbf{\bibinfo{volume}{27}}, \bibinfo{pages}{379 -- 423}
  (\bibinfo{year}{1948}).

\bibitem{shannonSeminal1}
\bibinfo{author}{Shannon, C.~E.}
\newblock \bibinfo{title}{{A Mathematical Theory of Communication}}.
\newblock \emph{\bibinfo{journal}{Bell Syst. Tech. J.}}
  \textbf{\bibinfo{volume}{27}}, \bibinfo{pages}{623--656}
  (\bibinfo{year}{1948}).

\bibitem{holevoBOOK}
\bibinfo{author}{Holevo, A.~S.}
\newblock \emph{\bibinfo{title}{{Quantum Systems, Channels, Information}}}
  (\bibinfo{publisher}{De Gruyter}, \bibinfo{address}{Berlin, Boston},
  \bibinfo{year}{2012}).
\newblock
  %\urlprefix\url{http://www.degruyter.com/view/books/9783110273403/9783110273403/9783110273403.xml}.

\bibitem{cavesDrum}
\bibinfo{author}{Caves, C.~M.} \& \bibinfo{author}{Drummond, P.~D.}
\newblock \bibinfo{title}{{Quantum limits on bosonic communication rates}}.
\newblock \emph{\bibinfo{journal}{Rev. Mod. Phys.}}
  \textbf{\bibinfo{volume}{66}}, \bibinfo{pages}{481--537}
  (\bibinfo{year}{1994}).

\bibitem{wildeBOOK}
\bibinfo{author}{Wilde, M.~M.}
\newblock \emph{\bibinfo{title}{{Quantum Information Theory}}}
  (\bibinfo{publisher}{Cambridge University Press},
  \bibinfo{address}{Cambridge}, \bibinfo{year}{2013}).
\newblock %\urlprefix\url{http://arxiv.org/abs/1106.1445 http://ebooks.cambridge.org/ref/id/CBO9781139525343}.
%\newblock arXiv:\eprint{1106.1445}.

\bibitem{hayashiBOOK}
\bibinfo{author}{Hayashi, M.}
\newblock \emph{\bibinfo{title}{{Quantum Information Theory}}}.
\newblock Graduate Texts in Physics (\bibinfo{publisher}{Springer Berlin
  Heidelberg}, \bibinfo{address}{Berlin, Heidelberg}, \bibinfo{year}{2017}).
\newblock %\urlprefix\url{http://link.springer.com/10.1007/978-3-662-49725-8}.

\bibitem{nChuangBOOK}
\bibinfo{author}{Nielsen, M.~A.} \& \bibinfo{author}{Chuang, I.~L.}
\newblock \emph{\bibinfo{title}{{Quantum Computation and Quantum Information}}}
  (\bibinfo{publisher}{Cambridge University Press},
  \bibinfo{address}{Cambridge}, \bibinfo{year}{2010}).
\newblock %\urlprefix\url{http://ebooks.cambridge.org/ref/id/CBO9780511976667}.

\bibitem{holevo1}
\bibinfo{author}{Holevo, A.~S.}
\newblock \bibinfo{title}{{Information-theoretical aspects of quantum
  measurement}}.
\newblock \emph{\bibinfo{journal}{Probl. Peredachi Informatsii}}
  \textbf{\bibinfo{volume}{9}}, \bibinfo{pages}{31--42} (\bibinfo{year}{1973}).

\bibitem{holevo2}
\bibinfo{author}{Holevo, A.}
\newblock \bibinfo{title}{{The capacity of the quantum channel with general
  signal states}}.
\newblock \emph{\bibinfo{journal}{IEEE Trans. Inf. Theory}}
  \textbf{\bibinfo{volume}{44}}, \bibinfo{pages}{269--273}
  (\bibinfo{year}{1998}).

\bibitem{holevo3}
\bibinfo{author}{Holevo, A.~S.}
\newblock \bibinfo{title}{{Quantum coding theorems}}.
\newblock \emph{\bibinfo{journal}{Russ. Math. Surv.}}
  \textbf{\bibinfo{volume}{53}}, \bibinfo{pages}{1295--1331}
  (\bibinfo{year}{1998}).
%\newblock arXiv:\eprint{9809023}.

\bibitem{schumawest1}
\bibinfo{author}{Schumacher, B.} \& \bibinfo{author}{Westmoreland, M.~D.}
\newblock \bibinfo{title}{{Sending classical information via noisy quantum
  channels}}.
\newblock \emph{\bibinfo{journal}{Phys. Rev. A}} \textbf{\bibinfo{volume}{56}},
  \bibinfo{pages}{131--138} (\bibinfo{year}{1997}).

\bibitem{schumawest2}
\bibinfo{author}{Hausladen, P.}, \bibinfo{author}{Jozsa, R.},
  \bibinfo{author}{Schumacher, B.}, \bibinfo{author}{Westmoreland, M.} \&
  \bibinfo{author}{Wootters, W.~K.}
\newblock \bibinfo{title}{{Classical information capacity of a quantum
  channel}}.
\newblock \emph{\bibinfo{journal}{Phys. Rev. A}} \textbf{\bibinfo{volume}{54}},
  \bibinfo{pages}{1869--1876} (\bibinfo{year}{1996}).

\bibitem{kingUnital}
\bibinfo{author}{King, C.}
\newblock \bibinfo{title}{{Additivity for unital qubit channels}}.
\newblock \emph{\bibinfo{journal}{J. Math. Phys.}}
  \textbf{\bibinfo{volume}{43}}, \bibinfo{pages}{4641--4653}
  (\bibinfo{year}{2002}).
%\newblock arXiv:\eprint{0103156}.

\bibitem{depolarizingCCap}
\bibinfo{author}{King, C.}
\newblock \bibinfo{title}{{The capacity of the quantum depolarizing channel}}.
\newblock \emph{\bibinfo{journal}{IEEE Trans. Inf. Theory}}
  \textbf{\bibinfo{volume}{49}}, \bibinfo{pages}{221--229}
  (\bibinfo{year}{2003}).
%\newblock arXiv:\eprint{0204172}.

\bibitem{gaussOpt}
\bibinfo{author}{Giovannetti, V.}, \bibinfo{author}{Holevo, A.~S.} \&
  \bibinfo{author}{Garc{\'{i}}a-Patr{\'{o}}n, R.}
\newblock \bibinfo{title}{{A Solution of Gaussian Optimizer Conjecture for
  Quantum Channels}}.
\newblock \emph{\bibinfo{journal}{Commun. Math. Phys.}}
  \textbf{\bibinfo{volume}{334}}, \bibinfo{pages}{1553--1571}
  (\bibinfo{year}{2015}).

\bibitem{gaussMaj}
\bibinfo{author}{Mari, A.}, \bibinfo{author}{Giovannetti, V.} \&
  \bibinfo{author}{Holevo, A.~S.}
\newblock \bibinfo{title}{{Quantum State Majorization at the Output of Bosonic
  Gaussian Channels}}.
\newblock \emph{\bibinfo{journal}{Nat. Commun.}} \textbf{\bibinfo{volume}{5}},
  \bibinfo{pages}{3826} (\bibinfo{year}{2014}).
%\newblock arXiv:\eprint{1312.3545}.

\bibitem{gaussCap}
\bibinfo{author}{Giovannetti, V.}, \bibinfo{author}{Garc{\'{i}}a-Patr{\'{o}}n,
  R.}, \bibinfo{author}{Cerf, N.~J.} \& \bibinfo{author}{Holevo, A.~S.}
\newblock \bibinfo{title}{{Ultimate classical communication rates of quantum
  optical channels}}.
\newblock \emph{\bibinfo{journal}{Nat. Photonics}}
  \textbf{\bibinfo{volume}{8}}, \bibinfo{pages}{796--800}
  (\bibinfo{year}{2014}).
%\newblock arXiv:\eprint{1312.6225}.

\bibitem{LOSSY}
\bibinfo{author}{Giovannetti, V.} \emph{et~al.}
\newblock \bibinfo{title}{{Classical Capacity of the Lossy Bosonic Channel: The
  Exact Solution}}.
\newblock \emph{\bibinfo{journal}{Phys. Rev. Lett.}}
  \textbf{\bibinfo{volume}{92}}, \bibinfo{pages}{027902}
  (\bibinfo{year}{2004}).
%\newblock arXiv:\eprint{0308012}.

\bibitem{amplitudeDamping}
\bibinfo{author}{Giovannetti, V.} \& \bibinfo{author}{Fazio, R.}
\newblock \bibinfo{title}{{Information-capacity description of spin-chain
  correlations}}.
\newblock \emph{\bibinfo{journal}{Phys. Rev. A - At. Mol. Opt. Phys.}}
  \textbf{\bibinfo{volume}{71}}, \bibinfo{pages}{1--12} (\bibinfo{year}{2005}).
%\newblock arXiv:\eprint{0405110}.

\bibitem{horodecki}
\bibinfo{author}{Horodecki, R.}, \bibinfo{author}{Horodecki, P.},
  \bibinfo{author}{Horodecki, M.} \& \bibinfo{author}{Horodecki, K.}
\newblock \bibinfo{title}{{Quantum entanglement}}.
\newblock \emph{\bibinfo{journal}{Quantum}} \bibinfo{pages}{110}
  (\bibinfo{year}{2007}).
%\newblock arXiv:\eprint{0702225}.

\bibitem{bennetEntAss1}
\bibinfo{author}{Bennett, C.~H.}, \bibinfo{author}{Shor, P.~W.},
  \bibinfo{author}{Smolin, J.~A.} \& \bibinfo{author}{Thapliyal, A.~V.}
\newblock \bibinfo{title}{{Entanglement-Assisted Classical Capacity of Noisy
  Quantum Channels}}.
\newblock \emph{\bibinfo{journal}{Phys. Rev. Lett.}}
  \textbf{\bibinfo{volume}{83}}, \bibinfo{pages}{3081--3084}
  (\bibinfo{year}{1999}).

\bibitem{bennetEntAss2}
\bibinfo{author}{Bennett, C.~H.}, \bibinfo{author}{Shor, P.~W.},
  \bibinfo{author}{Smolin, J.~A.} \& \bibinfo{author}{Thapliyal, A.~V.}
\newblock \bibinfo{title}{{Entanglement-assisted capacity of a quantum channel
  and the reverse Shannon theorem}}.
\newblock \emph{\bibinfo{journal}{IEEE Trans. Inf. Theory}}
  \textbf{\bibinfo{volume}{48}}, \bibinfo{pages}{2637--2655}
  (\bibinfo{year}{2002}).
%\newblock arXiv:\eprint{0106052}.

\bibitem{lloydQuant}
\bibinfo{author}{Lloyd, S.}
\newblock \bibinfo{title}{{Capacity of the noisy quantum channel}}.
\newblock \emph{\bibinfo{journal}{Phys. Rev. A}} \textbf{\bibinfo{volume}{55}},
  \bibinfo{pages}{1613--1622} (\bibinfo{year}{1997}).

\bibitem{shorQuantum}
\bibinfo{author}{Shor, P.~W.}
\newblock \bibinfo{title}{{The quantum channel capacity and coherent
  information}}.
\newblock In \emph{\bibinfo{booktitle}{Lect. Notes, MSRI Work. Quantum
  Comput.}} (\bibinfo{year}{2002}).

\bibitem{qCap1}
\bibinfo{author}{Barnum, H.}, \bibinfo{author}{Nielsen, M.~A.} \&
  \bibinfo{author}{Schumacher, B.}
\newblock \bibinfo{title}{{Information transmission through a noisy quantum
  channel}}.
\newblock \emph{\bibinfo{journal}{Phys. Rev. A}} \textbf{\bibinfo{volume}{57}},
  \bibinfo{pages}{4153--4175} (\bibinfo{year}{1998}).

\bibitem{qCap2}
\bibinfo{author}{Devetak, I.}
\newblock \bibinfo{title}{{The Private Classical Capacity and Quantum Capacity
  of a Quantum Channel}}.
\newblock \emph{\bibinfo{journal}{IEEE Trans. Inf. Theory}}
  \textbf{\bibinfo{volume}{51}}, \bibinfo{pages}{44--55}
  (\bibinfo{year}{2005}).

\bibitem{holGiovRev}
\bibinfo{author}{Holevo, A.~S.} \& \bibinfo{author}{Giovannetti, V.}
\newblock \bibinfo{title}{{Quantum channels and their entropic
  characteristics}}.
\newblock \emph{\bibinfo{journal}{Reports Prog. Phys.}}
  \textbf{\bibinfo{volume}{75}}, \bibinfo{pages}{046001}
  (\bibinfo{year}{2012}).

\bibitem{deg1}
\bibinfo{author}{Devetak, I.} \& \bibinfo{author}{Shor, P.~W.}
\newblock \bibinfo{title}{{The Capacity of a Quantum Channel for Simultaneous
  Transmission of Classical and Quantum Information}}.
\newblock \emph{\bibinfo{journal}{Commun. Math. Phys.}}
  \textbf{\bibinfo{volume}{256}}, \bibinfo{pages}{287--303}
  (\bibinfo{year}{2005}).

\bibitem{antiDeg}
\bibinfo{author}{Caruso, F.} \& \bibinfo{author}{Giovannetti, V.}
\newblock \bibinfo{title}{{Degradability of Bosonic Gaussian channels}}.
\newblock \emph{\bibinfo{journal}{Phys. Rev. A}} \textbf{\bibinfo{volume}{74}},
  \bibinfo{pages}{062307} (\bibinfo{year}{2006}).

\bibitem{deg2}
\bibinfo{author}{Cubitt, T.~S.}, \bibinfo{author}{Ruskai, M.~B.} \&
  \bibinfo{author}{Smith, G.}
\newblock \bibinfo{title}{{The structure of degradable quantum channels}}.
\newblock \emph{\bibinfo{journal}{J. Math. Phys.}}
  \textbf{\bibinfo{volume}{49}}, \bibinfo{pages}{1--42} (\bibinfo{year}{2008}).
%\newblock arXiv:\eprint{arXiv:0802.1360v2}.

\bibitem{weakDeg}
\bibinfo{author}{Caruso, F.}, \bibinfo{author}{Giovannetti, V.} \&
  \bibinfo{author}{Holevo, A.~S.}
\newblock \bibinfo{title}{{One-mode bosonic Gaussian channels: A full
  weak-degradability classification}}.
\newblock \emph{\bibinfo{journal}{New J. Phys.}} \textbf{\bibinfo{volume}{8}},
  \bibinfo{pages}{1--23} (\bibinfo{year}{2006}).
%\newblock arXiv:\eprint{0609013v2}.

\bibitem{DiVincenzo1998}
\bibinfo{author}{DiVincenzo, D.~P.}, \bibinfo{author}{Shor, P.~W.} \&
  \bibinfo{author}{Smolin, J.~A.}
\newblock \bibinfo{title}{{Quantum-channel capacity of very noisy channels}}.
\newblock \emph{\bibinfo{journal}{Phys. Rev. A - At. Mol. Opt. Phys.}}
  \textbf{\bibinfo{volume}{57}}, \bibinfo{pages}{830--839}
  (\bibinfo{year}{1998}).
%\newblock arXiv:\eprint{9706061}.

\bibitem{Smith2007}
\bibinfo{author}{Smith, G.} \& \bibinfo{author}{Smolin, J.~A.}
\newblock \bibinfo{title}{{Degenerate quantum codes for Pauli channels}}.
\newblock \emph{\bibinfo{journal}{Phys. Rev. Lett.}}
  \textbf{\bibinfo{volume}{98}} (\bibinfo{year}{2007}).
%\newblock arXiv:\eprint{0604107v2}.

\bibitem{zeroQCap}
\bibinfo{author}{Smith, G.} \& \bibinfo{author}{Yard, J.}
\newblock \bibinfo{title}{{Quantum Communication with Zero-Capacity Channels}}.
\newblock \emph{\bibinfo{journal}{Science}}
  \textbf{\bibinfo{volume}{321}}, \bibinfo{pages}{1812--1815}
  (\bibinfo{year}{2008}).
%\newblock arXiv:\eprint{0807.4935}.

\bibitem{hastings2009}
\bibinfo{author}{Hastings, M.~B.}
\newblock \bibinfo{title}{{Superadditivity of communication capacity using
  entangled inputs}}.
\newblock \emph{\bibinfo{journal}{Nat. Phys.}} \textbf{\bibinfo{volume}{5}},
  \bibinfo{pages}{255--257} (\bibinfo{year}{2009}).

\bibitem{supAddQCap}
\bibinfo{author}{Cubitt, T.} \emph{et~al.}
\newblock \bibinfo{title}{{Unbounded number of channel uses may be required to
  detect quantum capacity}}.
\newblock \emph{\bibinfo{journal}{Nat. Commun.}} \textbf{\bibinfo{volume}{6}},
  \bibinfo{pages}{6739} (\bibinfo{year}{2015}).

\bibitem{RevGauss}
\bibinfo{author}{Weedbrook, C.} \emph{et~al.}
\newblock \bibinfo{title}{{Gaussian quantum information}}.
\newblock \emph{\bibinfo{journal}{Rev. Mod. Phys.}}
  \textbf{\bibinfo{volume}{84}}, \bibinfo{pages}{621--669}
  (\bibinfo{year}{2012}).
%\newblock arXiv:\eprint{1110.3234}.

\bibitem{Xiang2017}
\bibinfo{author}{Xiang, Z.~L.}, \bibinfo{author}{Zhang, M.},
  \bibinfo{author}{Jiang, L.} \& \bibinfo{author}{Rabl, P.}
\newblock \bibinfo{title}{{Intracity quantum communication via thermal
  microwave networks}}.
\newblock \emph{\bibinfo{journal}{Phys. Rev. X}} \textbf{\bibinfo{volume}{7}},
  \bibinfo{pages}{011035} (\bibinfo{year}{2017}).
%\newblock arXiv:\eprint{1611.10241}.

\bibitem{Kurizki2015}
\bibinfo{author}{Kurizki, G.} \emph{et~al.}
\newblock \bibinfo{title}{{Quantum technologies with hybrid systems}}.
\newblock \emph{\bibinfo{journal}{Proc. Natl. Acad. Sci.}}
  \textbf{\bibinfo{volume}{112}}, \bibinfo{pages}{3866--3873}
  (\bibinfo{year}{2015}).
%\newblock arXiv:\eprint{1504.00158}.

\bibitem{qCapLossy}
\bibinfo{author}{Wolf, M.~M.}, \bibinfo{author}{P{\'{e}}rez-Garc{\'{i}}a, D.}
  \& \bibinfo{author}{Giedke, G.}
\newblock \bibinfo{title}{{Quantum Capacities of Bosonic Channels}}.
\newblock \emph{\bibinfo{journal}{Phys. Rev. Lett.}}
  \textbf{\bibinfo{volume}{98}}, \bibinfo{pages}{130501}
  (\bibinfo{year}{2007}).

\bibitem{holWer}
\bibinfo{author}{Holevo, A.~S.} \& \bibinfo{author}{Werner, R.~F.}
\newblock \bibinfo{title}{{Evaluating capacities of bosonic Gaussian
  channels}}.
\newblock \emph{\bibinfo{journal}{Phys. Rev. A - At. Mol. Opt. Phys.}}
  \textbf{\bibinfo{volume}{63}}, \bibinfo{pages}{1--14} (\bibinfo{year}{2001}).
%\newblock arXiv:\eprint{9912067}.

\bibitem{plob}
\bibinfo{author}{Pirandola, S.}, \bibinfo{author}{Laurenza, R.},
  \bibinfo{author}{Ottaviani, C.} \& \bibinfo{author}{Banchi, L.}
\newblock \bibinfo{title}{{Fundamental limits of repeaterless quantum
  communications}}.
\newblock \emph{\bibinfo{journal}{Nat. Commun.}} \textbf{\bibinfo{volume}{8}},
  \bibinfo{pages}{15043} (\bibinfo{year}{2017}).
%\newblock arXiv:\eprint{1510.08863}.

\bibitem{wildeDecDirettaBound}
\bibinfo{author}{Sharma, K.}, \bibinfo{author}{Wilde, M.~M.},
  \bibinfo{author}{Adhikari, S.} \& \bibinfo{author}{Takeoka, M.}
\newblock \bibinfo{title}{{Bounding the energy-constrained quantum and private
  capacities of phase-insensitive bosonic Gaussian channels}}.
\newblock \emph{\bibinfo{journal}{New J. Phys.}} \textbf{\bibinfo{volume}{20}},
  \bibinfo{pages}{063025} (\bibinfo{year}{2018}).
%\newblock arXiv:\eprint{1708.07257}.

\bibitem{deco2}
\bibinfo{author}{Garc{\'{i}}a-Patr{\'{o}}n, R.},
  \bibinfo{author}{Navarrete-Benlloch, C.}, \bibinfo{author}{Lloyd, S.},
  \bibinfo{author}{Shapiro, J.~H.} \& \bibinfo{author}{Cerf, N.~J.}
\newblock \bibinfo{title}{{Majorization Theory Approach to the Gaussian Channel
  Minimum Entropy Conjecture}}.
\newblock \emph{\bibinfo{journal}{Phys. Rev. Lett.}}
  \textbf{\bibinfo{volume}{108}}, \bibinfo{pages}{110505}
  (\bibinfo{year}{2012}).

\bibitem{serafiniBOOK}
\bibinfo{author}{Serafini, A.}
\newblock \emph{\bibinfo{title}{{Quantum Continuous Variables: A Primer of
  Theoretical Methods}}} (\bibinfo{publisher}{CRC Press},
  \bibinfo{year}{2017}).

\bibitem{cohInfo}
\bibinfo{author}{Schumacher, B.} \& \bibinfo{author}{Nielsen, M.~A.}
\newblock \bibinfo{title}{{Quantum data processing and error correction}}.
\newblock \emph{\bibinfo{journal}{Phys. Rev. A}} \textbf{\bibinfo{volume}{54}},
  \bibinfo{pages}{2629--2635} (\bibinfo{year}{1996}).

\bibitem{wilde2017}
\bibinfo{author}{Wilde, M.~M.}, \bibinfo{author}{Tomamichel, M.} \&
  \bibinfo{author}{Berta, M.}
\newblock \bibinfo{title}{{A meta-converse for private communication over
  quantum channels}}.
\newblock In \emph{\bibinfo{booktitle}{IEEE Trans. Inf. Theory}} \textbf{\bibinfo{volume}{63}},
  \bibinfo{pages}{1792--1817} (\bibinfo{publisher}{IEEE}, \bibinfo{year}{2017}).
%\newblock arXiv:\eprint{1602.08898}.

\bibitem{Christandl2017}
\bibinfo{author}{Christandl, M.} \& \bibinfo{author}{M{\"{u}}ller-Hermes, A.}
\newblock \bibinfo{title}{{Relative Entropy Bounds on Quantum, Private and
  Repeater Capacities}}.
\newblock \emph{\bibinfo{journal}{Commun. Math. Phys.}}
  \textbf{\bibinfo{volume}{353}}, \bibinfo{pages}{821--852}
  (\bibinfo{year}{2017}).
%\newblock arXiv:\eprint{1604.03448}.

\bibitem{Sutter2017}
\bibinfo{author}{Sutter, D.}, \bibinfo{author}{Scholz, V.~B.},
  \bibinfo{author}{Winter, A.} \& \bibinfo{author}{Renner, R.}
\newblock \bibinfo{title}{{Approximate Degradable Quantum Channels}}.
\newblock \emph{\bibinfo{journal}{IEEE Trans. Inf. Theory}}
  \textbf{\bibinfo{volume}{63}}, \bibinfo{pages}{7832--7844}
  (\bibinfo{year}{2017}).
%\newblock arXiv:\eprint{1412.0980}.

\bibitem{Yard2005}
\bibinfo{author}{Yard, J.}, \bibinfo{author}{Hayden, P.} \&
  \bibinfo{author}{Devetak, I.}
\newblock \bibinfo{title}{{Capacity theorems for quantum multiple-access
  channels: Classical-quantum and quantum-quantum capacity regions}}.
\newblock \emph{\bibinfo{journal}{IEEE Trans. Inf. Theory}}
  \textbf{\bibinfo{volume}{54}}, \bibinfo{pages}{3091--3113}
  (\bibinfo{year}{2008}).
%\newblock arXiv:\eprint{0501045}.

\bibitem{wolf2006}
\bibinfo{author}{Wolf, M.~M.}, \bibinfo{author}{Giedke, G.} \&
  \bibinfo{author}{Cirac, J.~I.}
\newblock \bibinfo{title}{{Extremality of Gaussian Quantum States}}.
\newblock \emph{\bibinfo{journal}{Phys. Rev. Lett.}}
  \textbf{\bibinfo{volume}{96}}, \bibinfo{pages}{080502}
  (\bibinfo{year}{2006}).

\bibitem{wildeThermOpt}
\bibinfo{author}{Davis, N.}, \bibinfo{author}{Shirokov, M.~E.} \&
  \bibinfo{author}{Wilde, M.~M.}
\newblock \bibinfo{title}{{Energy-constrained two-way assisted private and
  quantum capacities of quantum channels}}.
\newblock \emph{\bibinfo{journal}{Phys. Rev. A}} \textbf{\bibinfo{volume}{97}}, \bibinfo{pages}{062310}
  (\bibinfo{year}{2018}).
%\newblock arXiv:\eprint{1801.08102}.

\bibitem{holevo2008}
\bibinfo{author}{Holevo, A.~S.}
\newblock \bibinfo{title}{{Entanglement-breaking channels in infinite
  dimensions}}.
\newblock \emph{\bibinfo{journal}{Probl. Inf. Transm.}}
  \textbf{\bibinfo{volume}{44}}, \bibinfo{pages}{171--184}
  (\bibinfo{year}{2008}).
%\newblock arXiv:\eprint{arXiv:0802.0235v1}.

\end{thebibliography}

\section*{Author contributions}
M. Rosati, A. Mari and V. Giovannetti have all contributed equally to the research and to the writing of the article. 

\section*{Competing Interests}
The authors declare no competing interests.

\section*{Acknowledgements}
M.R. acknowledges funding by the Spanish MINECO, project FIS2016-80681-P with the support of AEI/FEDER funds, and the Generalitat de Catalunya, project CIRIT 2017-SGR-1127.

\end{document}